\documentclass[pdflatex,sn-mathphys-num]{sn-jnl}

\usepackage{graphicx}%
\usepackage{multirow}%
\usepackage{amsmath,amssymb,amsfonts}%
\usepackage{amsthm}%
\usepackage{mathrsfs}%
\usepackage[title]{appendix}%
\usepackage{xcolor}%
\usepackage{textcomp}%
\usepackage{manyfoot}%
\usepackage{booktabs}%
\usepackage{algorithm}%
\usepackage{algorithmicx}%
\usepackage{algpseudocode}%
\usepackage{listings}%

\theoremstyle{thmstyleone}%
\theoremstyle{thmstyletwo}%

\theoremstyle{thmstylethree}%

\begin{document}

\title{Challenges to the cosmological constant model following results from the Dark Energy Survey}


\author*[1]{\fnm{Santiago} \sur{Avila}}\email{santiagoj.avila@ciemat.es}

\author*[2]{\fnm{Juan} \sur{Mena-Fern\'andez}}\email{juanmenafdez95@gmail.com}

\author[3]{\fnm{Maria} \sur{Vincenzi}}

\affil[1]{Centro de Investigaciones Energ\'eticas, Medioambientales y Tecnol\'ogicas (CIEMAT), Madrid, Spain}

\affil[2]{Universit\'e Grenoble Alpes, CNRS, LPSC-IN2P3, 38000 Grenoble, France}

\affil[3]{Department of Physics, University of Oxford, Denys Wilkinson Building, Keble Road, Oxford OX1 3RH, United Kingdom}


\abstract{In the last year, several pieces of evidence have pointed to a possible deviation from the standard cosmological model, $\Lambda$CDM. The recent work by the Dark Energy Survey (DES) collaboration reports a preference in the ballpark of $3\sigma$ in favor of dynamical dark energy against the standard cosmological model. For that, it used its final analyses of Baryonic Acoustic Oscillations and type Ia Supernovae, both sensitive to the expansion history of the Universe, in combination with the Cosmic Microwave Background (CMB) from Planck. This adds to the growing debate about the nature of dark energy. \textit{Published as a Perspective in Nature Astronomy in August 2025}.}




\maketitle

\section{Coronation of $\Lambda$}\label{sec1}

At the turn of the century, the $\Lambda$CDM model was established as the concordance model. On the one hand, the cosmological constant, $\Lambda$, which was first introduced by Einstein to modify his equations describing general relativity \cite{Lorentz1923Principle}, was a natural explanation for the accelerated expansion of the Universe that had just been discovered in 1998 while studying type Ia Supernovae (SN). On the other hand, this same model was capable of reconciling the Cosmic Microwave Background (CMB) anisotropies and the observations of the Large-Scale Structure (LSS) \cite{Percival20012dF}. 

Type Ia Supernovae are standardized candles. This means that the rest-frame luminosity at the peak of their light curve –brightness as a function of time– can be predicted from the stretch of it. This allows cosmologists to establish the luminosity distance to these objects. If one measures the redshifts of SN or their host galaxies, the distance-redshift relation (known as the Hubble diagram) contains information on the expansion history of the Universe and the content of the Universe. In 1998, two groups measured for the first time the acceleration of the expansion of the Universe using these techniques \cite{Riess1998Observational,Perlmutter1999Measurements}.

Baryon acoustic oscillations (BAO) are imprints of sound waves that propagated in the primordial plasma of the early Universe. These oscillations froze at around the time of recombination, which left a characteristic preferred scale -the sound horizon at this epoch- in the distribution of galaxies that can be used as a standard ruler. By measuring the angle this scale subtends on the sky, cosmologists can, again, establish the relation between distance and redshift, rich in cosmological information. The first detection of the BAO feature in the clustering of galaxies was reported in 2005 by both the Sloan Digital Sky Survey (SDSS) \cite{Eisenstein2005Detection} and the 2dF Galaxy Redshift Survey (2dFGRS) \cite{Cole20052dF} collaborations.

During this epoch, several experiments (COBE \cite{Fixsen1996Cosmic}, BOOMERanG \cite{Netterfield2002Measurement}, WMAP \cite{Hinshaw2003FirstYear}) measured the power spectrum of the anisotropies of the CMB, increasing the area coverage and angular resolution over time. These measurements indicated the need for a flat universe, i.e., a universe that reaches the critical density. On the other hand, several LSS measurements had indicated that matter accounted for about 25\% of this critical density. Nevertheless, having a cosmological constant accounting for the remaining 75\% could reconcile all these data. The measurements of SN and the first detection of BAO confirmed this scenario, finding concordance among the different probes. By this time it was already well established that the majority of matter in the Universe was in the form of Cold Dark Matter (CDM), completing the $\Lambda$CDM concordance model \cite{PeeblesLarge,OstrikerSize,EinastoDynamic}. 

Whereas all these data were compatible in the $\Lambda$CDM model, other more complex scenarios could also fit them. In this context, the energy responsible for the accelerated expansion of the Universe was coined as dark energy, whether it was a cosmological constant or something more exotic. The advantage of $\Lambda$ is that it can be introduced in Einstein’s equations without modifying the fundamental symmetries and the dynamics they were intended to describe, but having an important impact on cosmology. Other simple physical models include quintessence –a dynamical scalar field minimally coupled to gravity, whose potential energy causes the accelerated expansion– and modifications of general relativity.

A useful way to quantify deviations from $\Lambda$CDM is through a series of parameters. For example, one could have $w$ as a free parameter describing the equation of state of dark energy, requiring $w<-1/3$ for an accelerating universe. Another popular parametrization is the so-called $w_0w_a$CDM model, which allows the equation of state to vary linearly with the scale factor of the Universe: $w(a)=w_0+w_a(1-a)$. The $w_0w_a$CDM model reduces to $\Lambda$CDM for $w_0=-1$ and $w_a=0$. 

To better understand the nature of dark energy, a series of experiments, informed by the Dark Energy Task Force \cite{Albrecht2006Report}, were designed to measure with higher and higher precision possible deviations from $\Lambda$CDM, recently crowned as the standard cosmological model. In this context, the Dark Energy Survey was envisioned as an experiment aiming to constrain the dark energy equation of state through multiple cosmological probes.

\section{The Dark Energy Survey}

The Dark Energy Survey (DES) \cite{DES2005Dark} was designed to study dark energy from four probes: the number of galaxy clusters as a function of redshift and mass, the weak gravitational lensing of galaxies, galaxy clustering with a focus on the BAO scale, and the Hubble diagram of type Ia SN. Combining them, DES aims to measure the equation of state parameter $w_0$ with a precision better than 5\%, and its time evolution $w_a$ with a precision of around 30\%.

For that, an international collaboration built the Dark Energy Camera (DECam) \cite{Flaugher2015DARK} with a 570 Mpix CCD imager. DECam was installed in the 4-meter Victor Blanco telescope in Chile with a system of five filters (g, r, i, z and Y) in the optical and near-infrared range. Over 6 years (between 2013 and 2019), DES surveyed approximately 5,000 square degrees of the southern sky (~1/8 of the full sky), mapping out the LSS of the Universe. In parallel, during the first five years, DES also repeatedly observed ten ~3 square-degree fields with a cadence of approximately 5 days to detect SN. The implications of two of its main probes, BAO and SN, will be discussed below.

\section{Toward the era of precision cosmology}

For a long period, $\Lambda$ happily reigned the cosmological model with the support of the majority of the community. As more experiments came into play, the majority would confirm the predictions of $\Lambda$CDM, but increase the precision of the parameters. In this context, the CMB experiment Planck released its main results between 2012 and 2018 \cite{AghanimPlanck}. Planck was a full-sky Cosmic Microwave Background experiment of the European Space Agency. Its exquisite data allowed the era of precision cosmology to become a reality. For example, within the flat-$\Lambda$CDM, it constrained the abundance of dark energy to $\Omega_\Lambda=1-\Omega_m=0.6847\pm0.0073$, reaching the goal of having 1\%-precision cosmology.
 
In these years, a few tensions and anomalies arose in some experiments, or when combining and comparing several of them. However, none of these tensions reached the level where all the following conditions held simultaneously: (a) the tension is sufficiently significant, (b) point to another preferred model, and (c) the tension is maintained by independent experiments.

\section{The annus horribilis of $\Lambda$}

Whereas the CMB can constrain the $\Lambda$CDM model very well, when opening the space for alternative late-time evolution of the Universe, it becomes less constraining and needs help from other probes such as SN or BAO to obtain well-constrained results. Before 2024, most of the combinations of CMB with BAO and SN were pointing to a recovery of the baseline $\Lambda$CDM. But this rapidly changed in this terrible year of $\Lambda$’s reign.

The Dark Energy Survey released its final supernova data in January 2024 \cite{Abbott2024DarkSN}. The most striking results showed up when allowing the equation of state of dark energy to vary with the Universe’s scale factor ($w_0w_a$CDM). DES SN on its own found a 2-$\sigma$ deviation from $\Lambda$CDM favouring $w_a<0$ and $w_0<-1$. That study also showed that the tension would significantly increase when adding more data from BAO, CMB and the combination of galaxy clustering and weak lensing.

One month later, the Dark Energy Survey released its final analysis of the Baryonic Acoustic Oscillations \cite{Abbott2024DarkBAO}. This allowed DES to set an angular distance measurement at redshift $z\sim0.85$, with a precision of $2.1\%$. The measurement was found to be 4.3\% below Planck-$\Lambda$CDM prediction, showing a $\sim 2\sigma$ deviation.

In April 2024, the Dark Energy Spectroscopic Instrument (DESI) released its first cosmological results based on the precise measurements of the BAO feature in seven redshift bins \cite{Adame2025DESI}. Whereas DESI alone did not show a significant deviation from $\Lambda$CDM, when combined with Planck CMB and different SN datasets showed a preference for $w_0w_a$CDM compared to $\Lambda$CDM ranging from 2.5 to 3.9$\sigma$, depending on the SN dataset. The most significant deviation occurred when considering the DES SN in the analysis.

All of a sudden, the reign of $\Lambda$ could be about to fall, as increasing evidence for evolving dark energy appeared in the scene.

\section{Implications for dark energy from the DES expansion history probes}

\begin{figure}
    \centering
    \includegraphics[width=\linewidth]{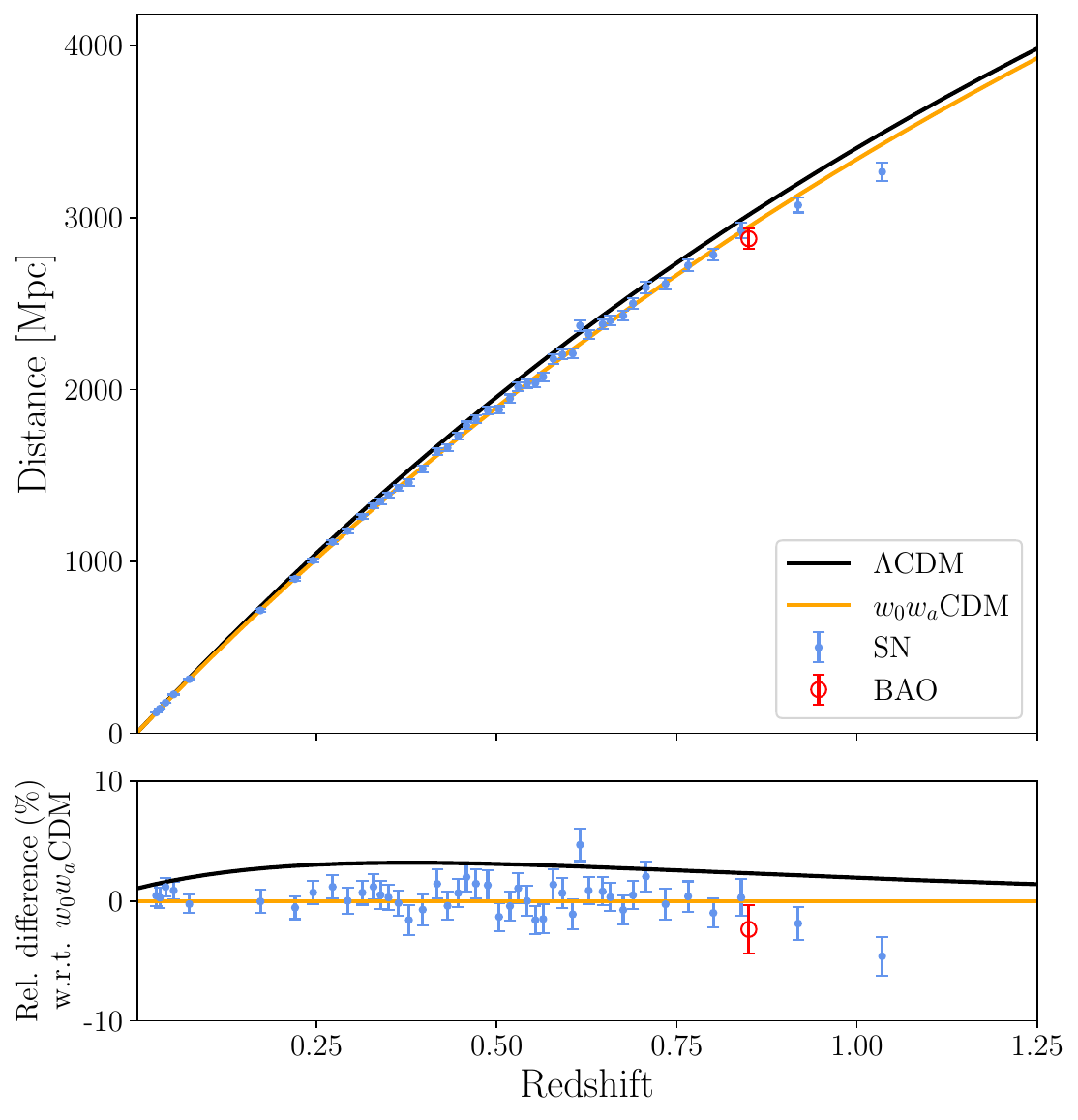}
    \caption{Measurements of distances from the DES expansion history probes. We include the results for type Ia Supernovae (blue) and BAO (red) with error bars that account for both statistical and systematic uncertainties. Data are compared with the predictions of a universe where dark energy is the cosmological constant (black line) and with a universe where dark energy changes with time (orange line). The top panel represents the distance-to-redshift relation, and the bottom one shows the relative differences (in percentage) with respect to the best-fit $\Lambda$CDM model (black) and $w_0w_a$CDM model (orange). Data are better fitted by the orange line.
    Credits: Dark Energy Survey. Figure adapted from \cite{Collaboration2025Dark}.}
    \label{fig:hubble}
\end{figure}

The latest DES BAO and type Ia SN results are the culmination of years of collaborative work and represent what the completed DES can tell us about the expansion of the Universe via geometric probes. The DES BAO and SN distance-to-redshift measurements are shown in \autoref{fig:hubble}, and the implications for the cosmological model are explored in a recent paper \cite{Collaboration2025Dark}.

The old concordance model, $\Lambda$CDM, is not able to reconcile DES observables combined with external datasets, finding significant tensions. Additionally, DES BAO and SN prefer, respectively, lower and higher values of the abundance of matter than CMB. There is no combination of parameters within $\Lambda$CDM that can explain DES BAO, DES SN and Planck CMB simultaneously.

One-parameter extensions, such as allowing a free parameter for the curvature of the Universe or a model with a free parameter for the equation of state of dark energy ($w$), are not sufficient to reconcile our observables. In the latter, BAO and SN again push in different directions of the parameter space, favoring $w<-1$ and $w>-1$, respectively. 
The inability to successfully interpret the different datasets within the same model framework, added to the fact that BAO and SN do not probe the same redshift ranges, led the collaboration to test the $w_0w_a$CDM model, in which the equation of state of dark energy is allowed to vary with time. Interestingly, all datasets tested become compatible when interpreted in $w_0w_a$CDM, and DES finds that the fit to $w_0w_a$CDM is significantly better compared to $\Lambda$CDM.

\begin{figure}
    \centering
    \includegraphics[width=\linewidth]{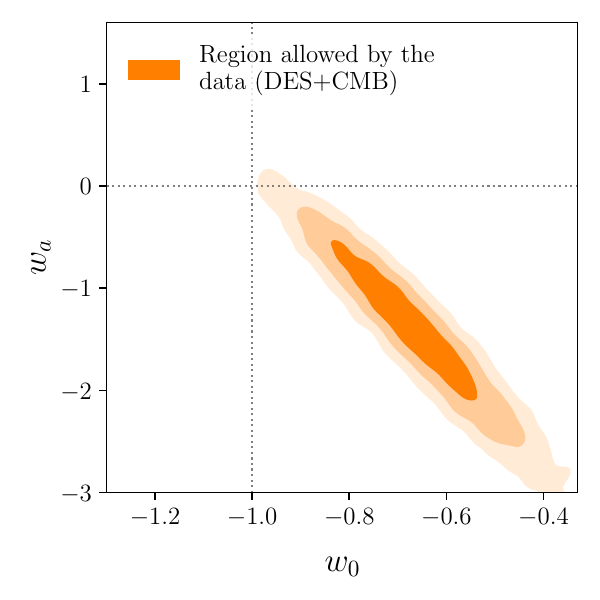}
    \caption{Determination of the parameters of the dark energy from DES geometric probes (BAO and SN) and the Cosmic Microwave Background (CMB) from Planck. The orange contours (from darker to lighter) represent the regions enclosing 68.3\%, 95.6\% and 99.7\% (1, 2 and 3$\sigma$, respectively) of the total probability after marginalizing over the other parameters of the model. A cosmological constant would be in the intersection of the vertical and horizontal lines (-1,0), However, data favor values inside the orange region, where dark energy changes with time.
    Credits: Dark Energy Survey. Figure adapted from \cite{Collaboration2025Dark}.}
    \label{fig:w0wa}
\end{figure}

The baseline result is shown in \autoref{fig:w0wa}, displaying the constraints in $w_0$ and $w_a$ from the combination of DES BAO, DES SN and Planck CMB. The constraint reads
$w_0=-0.673^{+0.098}_{-0.097}, w_a=-1.37^{+0.51}_{-0.50}$,
where the errors are the asymmetric 68.3\% credible intervals, showing a $3.2\sigma$ deviation from $\Lambda$CDM, which is given by $w_0=-1$ and $w_a=0$. From a more frequentist approach, an improvement in the goodness of fit of  $\Delta \chi^2=$11.6 for only 2 extra parameters is found, also very significant.   

When trying different data combinations, they all agree in the $w_0>-1$ and $w_a<0$ quadrant, finding a good consistency among all datasets considered. This is also consistent with the results preferred by DESI BAO in combination with other datasets \cite{Adame2025DESI}. DESI and DES only overlap at a small area and redshift range, and therefore there is a small correlation between the BAO measurements of the two surveys. If this correlation was neglected, the combination of the DES final BAO and the DESI BAO from the first year of data together with DES SN and CMB would reach a 4.2$\sigma$ deviation from $\Lambda$CDM. Very recently, DESI also reported a similar level of tension using its BAO from its first three years of data, in combination with CMB and DES SN \cite{Collaboration2025DESI}. 

In this context, it is remarkable that one single experiment, DES, can obtain competitive SN and BAO results to understand the late-time evolution of the Universe and complement the early-time information from the CMB. We are now in a moment where not only one BAO, one SN and one CMB experiment is showing significant deviation from $\Lambda$CDM, but the community has provided redundancy in each of the datasets. If in the coming years the significance of the deviation increases, we could find ourselves in an interesting scenario where we would rule out $\Lambda$CDM. If the reign of $\Lambda$ falls, who will take the throne of dark energy?

\section{If $\Lambda$ were to fall}

There is growing evidence for an expansion history of the Universe beyond $\Lambda$CDM. This is best pictured by a strong preference for the $w_0w_a$CDM model compared to $\Lambda$CDM, in the ballpark of 3 or 4$\sigma$. Unless unknown systematics were to affect more than one dataset, the simplest explanation for the data is that dark energy could be more exotic than the cosmological constant, i.e., its energy density could be evolving with time. If confirmed, this would shake the foundations of fundamental physics.

As explained earlier, the cosmological constant had a very natural explanation as a minimal modification of Einstein’s equations without introducing new physics. Nevertheless, the value we measure today of $\Lambda$ poses questions about why it is so small compared to the predictions of vacuum energy from elementary physics and why its energy density is comparable to that of matter today. Answering these questions becomes a somewhat speculative matter that involves pushing our fundamental theories beyond their tested range of validity, bringing together gravitation and particle physics. One popular but controversial solution invokes the anthropic principle: we could live in a bubble universe that has a particularly low value of $\Lambda$, as much larger values would stall structure formation even at proto-galactic scales. This way, humanity and cosmology could only develop in those bubble universes with low enough $\Lambda$ regardless of how rare these bubbles are \cite{Weinberg1989cosmological}. Additionally, the fact that matter and $\Lambda$ have similar density today may not be that rare after all: the Universe has spent a quarter of its lifetime after $\Lambda$-matter equality and it did not have enough time before this equality to form the astrophysical environment that we observe \cite{Velten2014Aspects}.

On the other hand, if the explanation of accelerated expansion is not given from a cosmological constant, this would imply new fundamental physics. This could be a non-minimal modification of Einstein’s general relativity equations, it could potentially be a new field beyond the standard model of particle physics or that some of the known components of the Universe behave in a non-standard manner. Alternatively, it could be a violation of one of the basic assumptions that build the $\Lambda$CDM model.

The $w_0w_a$CDM parametrization is a handy expansion to look for deviations from $\Lambda$CDM or a model with a constant equation of state or dark energy, but it is also able to match the predictions for the expansion history of the Universe in a wide range of physically-motivated models \cite{Chevallier2001ACCELERATING,Linder2003Exploring,de2008Calibrating}. This includes models with thawing behavior, in which dark energy behaves as a cosmological constant at early times and then $w(a)$ increases as the Universe expands (e.g., the pseudo-Nambu Goldstone model \cite{Frieman1995Cosmology}), and models with freezing behavior, in which dark energy has $w(a)>-1$ at early times and then it evolves to a cosmological constant at late times (e.g., supergravity models, which combine the principles of supersymmetry and general relativity \cite{Brax1999Quintessence}).

The recent DES results suggest that $w_0>-1$ and $w_a<0$. Taking these at face value with $w(a)=w_0+(1-w_a)a$ has important theoretical implications, since it hints at a \textit{phantom} behavior of dark energy in the \textit{recent} past, at $z\sim0.3$. This phantom behavior arises whenever $w(a)<-1$, which leads to an increase in the energy density of dark energy as the Universe expands that can be seen from the continuity equation:
\begin{equation}
    \dot{\rho}_{\rm DE}=-3H(\rho_{\rm DE}+p_{\rm DE})=-3H\rho_{DE}(1+w(a))>0 \to w(a)<-1.\nonumber
\end{equation}
This violates the null energy condition, which requires that the energy density decreases with the expansion of the Universe. As a result, the stability of the theory is no longer guaranteed \cite{Caldwell2002phantom}. The phantom behavior represents serious theoretical difficulties for single scalar-field models of dark energy \cite{Carroll2003Can}, but modified gravity or multi-field models of dark energy can evade these difficulties \cite{Hu2005Crossing}.

The preference for the $w_0>-1$ and $w_a<0$ region would also suggest that dark energy is weakening with time: the accelerated expansion of the Universe might come to an end at some point in the future. This can have interesting consequences if someone wonders about the fate of the Universe or theorizes about futuristic intergalactic travel or communication. If the Universe continued to accelerate as in $\Lambda$CDM, at some point the comoving event horizon would get close enough to us that any galaxy outside our local group would be invisible and out of reach. On the other hand, if the acceleration were to stop, this would not happen, and we could theoretically travel or send signals to other galaxies outside our local group, even in the far future \cite{Rindler1956Visual}.

Eventually, if evidence ends up definitively ruling out $\Lambda$CDM, the community would need to focus on pinning down what physical model –among or beyond the examples listed in the discussion above– is the most viable to explain the expansion history of the Universe and determine what its fate will be. Interpreting the phantom crossing or the eventual fate of the Universe should be taken with care, as we could simply be extrapolating a simple parametrization, $w(a)=w_0+w_a(1-a)$, beyond its range of validity.

\section{Toward the next battle}

Current winds blow against a scenario where $\Lambda$ could maintain the crown much longer. The deviations from $\Lambda$CDM lie in the ballpark of 3 to 4 $\sigma$.\footnote{Since the publication of this perspective, an improved calibration of the DES SN appeared \cite{Dovekie}, slightly lowering the deviation to $\sim$2.5$\sigma$ (with DES BAO and CMB \cite{Mena26}) or $\sim3\sigma$ (with  DESI BAO and CMB \cite{Dovekie}).} Hence, the golden rule of 5 $\sigma$ significance (corresponding to a 1-in-a-million chance of being a statistical fluke) to claim a discovery could be at reach very soon. Nevertheless, more evidence is needed before the reign of $\Lambda$ can be turned down and new evidence could possibly weaken existing tensions, as seen with other tensions formerly reported in cosmology.

Other observables that can give us more insight into the nature of dark energy are related to how inhomogeneities grow in the expanding Universe. These could help distinguish scenarios of dark energy that behave identically when only looking at the expansion history. The Dark Energy Survey will be soon providing the state-of-the-art growth of structure constraints from its final dataset.\footnote{Some results appeared in \cite{DESY63x2pt} and the final $w_0-w_a$ constraints are in preparation.} This will include studies of weak gravitational lensing from exquisite measurements of the shape of over one hundred million galaxies \cite{Yamamoto2025Dark}, the clustering of galaxies and galaxy cluster counts as a function of mass and redshift \cite{To2025Dark}. Additionally, new generation dark energy experiments, such as DESI, Euclid \cite{Mellier2025Euclid} or the Vera C. Rubin’s LSST \cite{Ivezic2019LSST} will provide new data in the near future. If any of these further supported the deviation from $\Lambda$CDM, they could be the last straw for $\Lambda$’s reign, reaching the 5$\sigma$ barrier when combined with existing data.

If that threshold is met, we could find ourselves in a scenario with (a) a 1-in-a-million significance deviation from $\Lambda$CDM, (b) favoring an alternative model ($w_0w_a$CDM, which maps well to more physically-motivated models), and (c) where the community has shown consistency across different datasets. Whether we arrive at this scenario or the evidence goes away needs to be determined by the data, with exhaustive redundancies and checks for unexpected systematic errors. Meanwhile, the community is getting ready for the \textit{final battle} that could change the model that has governed cosmology over the past quarter of a century.

\section*{Acknowledgments}

SA has been supported by the Ramon y Cajal fellowship (RYC2022-037311-I) funded by the State Research Agency of the Spanish Ministry of Science and Innovation (MCIN/AEI/10.13039/501100011033) and Social European Funds plus (FSE+). The authors are grateful for having been members of the Dark Energy Survey collaboration, allowing them to participate in the analyses that led to the results discussed in this paper. The insights presented in this article represent only the views of the authors.

\section*{Author contributions}

S.A., J.M.F. and M.V. coordinated the preparation and defined the structure and main message of the manuscript. S.A. and J.M.F. wrote the body of the article. S.A., J.M.F. and M.V. were heavily involved in the science analysis that preceded this Perspective. J.M.F. prepared the figures.

\section*{Competing interests}
The authors declare no competing interests.

\bibliography{sn-bibliography}

\end{document}